# Compressed Sensing for Reconstructing Coherent Multidimensional Spectra

Zhengjun Wang[†], Shiwen Lei[‡,#], Khadga Jung Karki[†], Andreas Jakobsson[‡] and Tönu Pullerits[†]

[†] Division of Chemistry Physics and NanoLund, Lund University, P.O. Box 124, 22100, Lund, Sweden.
[‡] Centre of Mathematical Sciences, Lund University, P.O. Box 118, 22100, Lund, Sweden
[#] University of Electronic Science and Technology of China, 611731, Chengdu, China

## ABSTRACT

We apply two sparse reconstruction techniques, the least absolute shrinkage and selection operator (LASSO) and the sparse exponential mode analysis (SEMA), to two-dimensional (2D) spectroscopy. The algorithms are first tested on model data, showing that both are able to reconstruct the spectra using only a fraction of the data required by the traditional Fourier-based estimator. Through the analysis of a sparsely sampled experimental fluorescence detected 2D spectra of LH2 complexes, we conclude that both SEMA and LASSO can be used to significantly reduce the required data, still allowing to reconstruct the multidimensional spectra. Of the two techniques, it is shown that SEMA offers preferable performance, providing more accurate estimation of the spectral line widths and their positions. Furthermore, SEMA allows for off-grid components, enabling the use of a much smaller dictionary than the LASSO, thereby improving both the performance and lowering the computational complexity for reconstructing coherent multidimensional spectra.

## INTRODUCTION

Coherent multidimensional spectroscopy[1] has become an important technique for studying excited state dynamics in complex systems with congested spectral bands. It has been successfully applied in systems such as light harvesting complexes,[2,3] quantum dots,[4,5] quantum wells,[6] molecular aggregates[7,8,9] and more. In conventional photon echo based 2D spectroscopy, only the so-called coherence and population times are scanned, while the signal is recorded using a spectrometer directly providing spectral dependence of the detection without the need for explicit scanning of the corresponding time delay.[10] In recent developments, the coherent 2D spectroscopy is detected via various incoherent "action" signals. Fluorescence,[11] photocurrent,[12] photoelectron[13], and photoion[14] detection has been used so far. In these experiments, four laser pulses are used, which means that three time delays between the pulses need to be explicitly scanned. This can make multidimensional spectroscopy experiment very time-consuming. In such experiments, efficient data acquisition algorithms become essential. One promising approach is to use dictionary-based sparse reconstruction techniques in a compressed sensing context, such as the least absolute shrinkage and selection operator (LASSO) introduced by Tibshirani.[15] By including a penalty in the cost function, such techniques may be used to reconstruct non-uniformly sampled data sets that are well detailed using only a few components.[16] The technique has recently been applied to a variety of different spectroscopy experiments, for instance, X-ray diffraction,[17] 2D infrared spectroscopy,[18] multidimensional nuclear magnetic resonance,[19] atomic force microscopy,[20] mass spectrometry,[21] and coherent 2D spectroscopy.[22] Here, we examine the reconstruction of the fluorescence detected coherent two-dimensional (FD2D) spectra using two sparse reconstruction techniques, namely the aforementioned LASSO, and the recent Sparse Exponential Mode Analysis (SEMA) method.[23,24] We apply the method to the experimental data of the peripheral light harvesting antenna complexes (LH2) of photosynthetic purple bacteria shown in Figure 1 (leftmost).[25,26,27] The LH2 consists of two rings of bacteriochlorophyll (BChl) molecules, called B800 and B850.[28] In many purple bacteria, the B800 ring contains nine well-separated BChl molecules with an absorption band at about 800 nm, while the B850 ring has eighteen closely packed BChl molecules absorbing around 850 nm.[29]

A typical FD2D spectrum of such system with two clear linear absorption bands has four peaks.[30, 31, 32] In Figure 1 (rightmost), a 2D spectrum of LH2 is shown. The four peaks R11, R12, R21, and R22 can be clearly distinguished.

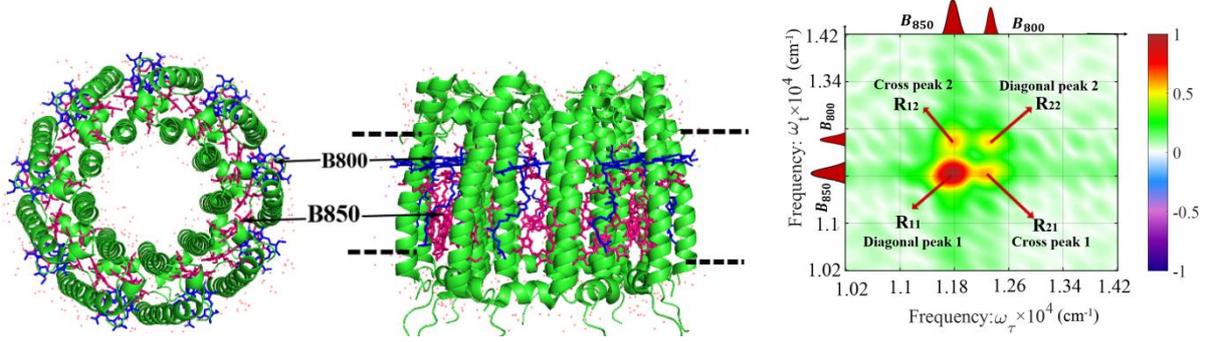

**Figure 1.** Left: the structure of LH2 complex. Right: FD2D spectrum of LH2. See text for details.

For constructing FD2D spectra, the fluorescence intensity is recorded while scanning four femtosecond laser pulses in respect of each other. In order to separate the correct nonlinear fluorescence signal due to the interaction of all four pulses from other possible signals (for example fluorescence excited by a single pulse), we use phase modulation technique together with smart lock-in type demodulation[33, 34]. Let τ, T, and t denote the time delays between the first and second pulses, the second and third pulses, and the third and fourth pulses, respectively. These times are also known as the coherence time, the population time, and the detection time. In our data set, we have recorded 20 points for T (from 0fs to 73fs) and 40 points for both τ and t. Taking Fourier transform over τ and t, yields a 2D spectrum for each T.

In this article, we apply sparse sampling to reduce the time it takes to record a 2D spectrum and investigate how noise influences reconstruction accuracy. Previously, sparse reconstruction techniques were examined by Roeding et al, who employed a two-step iterative shrinkage/thresholding (TwIST) algorithm, showing that spectra could be accurately reconstructed using only 25% of full data set.[35] Similar results were shown by Sanders et al who used a matching pursuit algorithm to reconstruct the data from atomic Rb vapour.[36] Hutson et al reconstructed the spectra using non-uniformly sampled data, using the projection-slice theorem on the multidimensional coherent spectrum.[37] The results of spectral band (damping) and the spectral width (frequency) were not analysed, only the spectra were sparsely reconstructed by the noted methods. In this work, we compare the reconstruction of sparsely sampled LH2 spectra by LASSO and the recently developed SEMA method[24].

**METHODS**
In order to formulate the LASSO and SEMA estimators, let

$$s = \left[ t^{(1)}_{i_1}, t^{(2)}_{i_2} \right]^T \quad (1)$$

denote the 2-D sampling times of the signal, with $t^{(1)}_{i_1}$ and $t^{(2)}_{i_2}$ denoting the $i_1$-th sampling point in the first dimension and $i_2$-th sampling point in the second dimension. These two dimensions correspond to τ and t in above time delay notation. In general, these sampling times may be arbitrarily selected in both dimensions, creating a non-uniform sampling grid. In the experiment analyzed here, the signal was sampled uniformly at 40 sampling points in these two dimensions. The signal may thus be represented as

$$x(s) = \sum_{k=1}^{K} g_k e^{\left[iw_k^{(1)} - \beta_k^{(1)}\right]t_{i_1}^{(1)}} \cdot e^{\left[iw_k^{(2)} - \beta_k^{(2)}\right]t_{i_2}^{(2)}} + \mu(s) \qquad (2)$$

with $w_k^{(l)}$ and $\beta_k^{(l)}$ denoting the frequency and width (damping) in the *l*-th dimension of the *k*-th 2D spectral band, where *l* = 1 or 2. Thus, we use here Lorentzian lineshape model. We point out that one may extend the model to include more detailed lineshapes, such as, for example Voigt model. More complex lineshapes typically come with the cost of more parameters and would lead to larger dictionary space. In the interest of brevity, we here limit the discussion to the Lorentzian, referring the interested reader to ref paper[38] for a further discussion on more detailed lineshapes.

In the above example of LH2 with two 1 D spectral bands, there are $K = 4$ possible 2D bands. Furthermore, $g_k$ denotes the complex numbered amplitude of the spectral band *k* where the imaginary part gives the initial phase which is here taken zero. The noise term $\mu(s)$ is assumed to be well modelled as Gaussian distributed random numbers. The noise amplitude (FWHM of the Gaussian distribution) is set to be 100 times lower than the amplitude of the spectral bands $g_k$. In the following, we use eq. (2) to construct model data sets which resemble the 2D spectroscopy experiment in order to test the efficiency of the LASSO and SEMA methods to recover the parameters $w_k^{(l)}, \beta_k^{(l)}$ from the data sets with different density of sampling down to just a few percent of the original number of points.

In order to recover the input parameters $w_k^{(l)}, \beta_k^{(l)}$, a so-called dictionary is formed over a sufficiently extensive set of possible values of the 2D band frequencies and damping constants $w_{p_1}^{(1)}, w_{p_2}^{(2)}, \beta_{j_1}^{(1)}$, and $\beta_{j_2}^{(2)}$. This allows the sum in eq. (2) to be extended to contain $m = 1,..., P_1 \times P_2 \times J_1 \times J_2$ terms, with $P_1, P_2, J_1,$ and $J_2$ denoting the number of frequency and damping dictionary elements in the two dimensions and $\tilde{g}_m$ giving the corresponding amplitude of the spectral component. One may then determine the parameters describing the signal by determining the non-zero components best fitting the penalized minimization problem:

$$\underset{\tilde{g}}{\text{minimize}} \left\{ \left\| vec\{x(s)\} - \tilde{A}\tilde{g} \right\|_2^2 + \sum_{m=1}^{P_1 \times P_2 \times J_1 \times J_2} \lambda |\tilde{g}_m| \right\} \qquad (3)$$

where $\lambda$ is a regularization coefficient which adds a penalty (we use $\lambda = 0.4$), $\tilde{g}$ is an amplitude vector formed from the vectorization, and the spectral bands $\tilde{A}$ are constructed from all possible candidates in the dictionary:

$$\tilde{A} = vec \left\{ e^{\left[iw_{p_1}^{(1)} - \beta_{j_1}^{(1)}\right]t_{i_1}^{(1)}} \circ e^{\left[iw_{p_2}^{(2)} - \beta_{j_2}^{(2)}\right]t_{i_2}^{(2)}} \right\} \qquad (4)$$

with $\circ$ denoting the outer product. As the result of the optimization all the spectral components not coinciding with the terms in eq. (2) will have very low amplitudes. The added penalty ensures that a solution ideally contains only the sought terms (others have negligible amplitude), allowing the corresponding terms to be identified by the components with largest amplitudes. Regrettably, even for a very coarse grid, the dimensionality of this minimization is computationally prohibitive, and the LASSO solution can in practice only be obtained by removing the influence of the damping components, setting $\beta_k^{(l)} = 0$. This allows the LASSO to determine the sought frequencies; these may then be used to simulate the signal for the missing sampling times such that one constructs a reconstructed data set over a uniformly sampled grid. From this, a 2D spectrum is then estimated using the fast Fourier method, from which the $\beta$ component may be estimated as the resulting line width of the peaks, at the determined frequencies. The SEMA estimator, on the other hand, introduces an iterative dictionary-learning step allowing the $\beta$ components to be incorporated without increasing the dimensionality of the minimization. This is done by initially assuming no damping for any of the used spectral components; then after first determining the suitable frequencies, the spectral components are updated to include a least-squares estimated damping (linewidth). The spectral components are further refined within narrow regions of the above suitable frequencies. Thereafter the fitting procedure is iterated, to further refine the estimates, first along frequency, and then over the damping parameter. We point out that although this implies that the initial fitting assumes a certain signal model (here Lorentzian), the found frequencies can be shown to still be accurate, despite the possible model mismatch[38]. Consequently, the parameters can be estimated without reconstructing the full data set. We refer the reader to SI for more thorough discussion of the compression algorithms and to references [24] for further details on the SEMA algorithm.

**RESULTS AND DISCUSSION**

In order to compare the ability of the two discussed methods to determine the sought parameters, we have generated P=2000 uniformly sampled Monte-Carlo simulations containing four spectral components, in which the frequencies were each drawn uniformly over [0.1, 0.97] and dampings was each drawn uniformly over [0.019, 0.035]. For each simulation, the signals were then subsampled at uniformly distributed time locations to yield the expected non-uniformly sampled data sets. We note that a suitable selection of samples will allow for improved estimation performance[23, 37]. Here, for simplicity, and as we mainly wish to illustrate the performance difference between the algorithms for a given set of samples, we use a random sampling scheme. In each simulation, the signal was corrupted by an additive Gaussian noise. The parameters of the four components were then estimated for each simulation, using 256 dictionary elements for each parameter $w_{p_1}^{(1)}, w_{p_2}^{(2)}, \beta_{j_1}^{(1)}$, and $\beta_{j_2}^{(2)}$. It should be stressed that these dictionary elements will most likely not coincide with the simulated parameter, to mimic the situation one may expect in a real experiment. Figures 2 and 3 show the resulting averaged root mean squared error (RMSE) of the frequency and damping parameters, respectively, when retaining varying degrees of randomly selected data points. Here, the RMSE of the frequency parameters has been computed as

$$RMSE_{frequency} = \sqrt{\frac{1}{8}\frac{1}{P}\sum_{p=1}^{P}\sum_{k=1}^{4}\sum_{l=1}^{2}(\frac{w_{k,(p)}^{(l)} - \hat{w}_{k,(p)}^{(l)}}{w_{k,(p)}^{(l)}})^2} \qquad (5)$$

where $w_{k,(p)}^{(l)}$ and $\hat{w}_{k,(p)}^{(l)}$ denote the true and the recovered frequency of the p[th] simulation for the k[th] spectral band, in dimension *l*, for *l*=1 or 2. The RMSE of the damping parameters is constructed similarly, as

$$RMSE_{damping} = \sqrt{\frac{1}{8}\frac{1}{P}\sum_{p=1}^{P}\sum_{k=1}^{4}\sum_{l=1}^{2}(\frac{\beta_{k,(p)}^{(l)} - \hat{\beta}_{k,(p)}^{(l)}}{\beta_{k,(p)}^{(l)}})^2} \qquad (6)$$

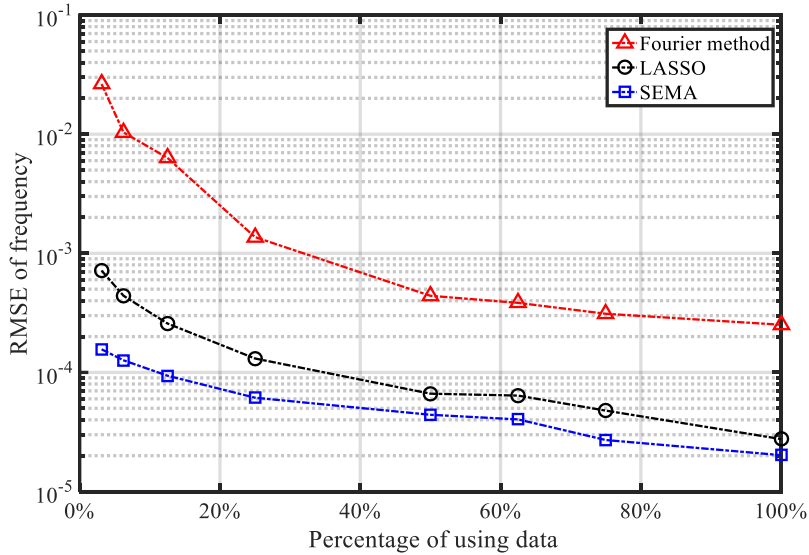

**Figure 2.** The summed RMSE of the frequency estimates.

As shown in Figure 2, the RMSE of the frequency parameters obtained from the discrete Fourier transform (Fourier method), which is computed as the peak values of the magnitude of the Fourier method of the data, as well as the LASSO and SEMA methods, decreases as the number of sampling point's increases. As is clear from the figure, the sparse reconstruction techniques are able to achieve significantly better performance than the Fourier method estimator, with SEMA showing the best performance. For the damping parameters, the LASSO is first used to reconstruct a uniformly sampled data set, from which the spectrum is computed using the Fourier method. From this the dampings are then estimated as the full width of half the maximum value for the found peak frequencies. For the Fourier method, the damping estimates are instead formed as the full width of half the maximum value of the peak of the magnitude of the Fourier method of the (non-uniformly sampled) data set, whereas the SEMA algorithm directly estimates the damping parameters, without reconstructing any uniformly sampled data set.

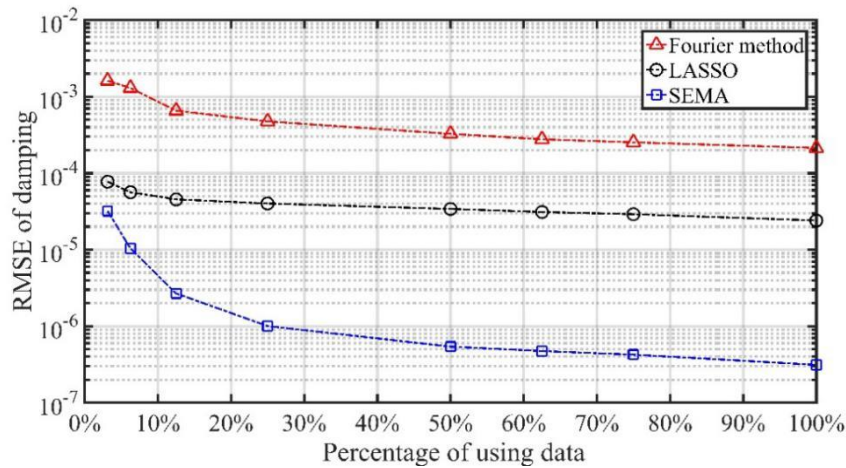

**Figure 3.** The summed RMSE of the damping estimates.

For both the frequency and damping estimates, the SEMA method is found to substantially outperform the LASSO and the Fourier method approach. The reason for this improvement is that SEMA forms a sparse estimate of both the frequencies and the dampings directly, while also allowing for off-grid frequencies, instead of basing the estimates on the reconstructed data, as the LASSO does. The poor estimate of the Fourier method estimate results from the sidelobes and spurious peaks resulting from computing the spectral estimate from a small non-uniformly sampled data set.

In time domain, most of the useful information is concentrated to the lower left corner of the plot with τ and t less than 100 fs, see Fig 4. The rest of the data correspond to longer times where the valuable spectral features have dephased or decayed away. The 2D spectra of B800 and B850 can be successfully reconstructed using only a fraction of the full data set (as is illustrated in Figure 5, showing 2D spectral estimates at T=70fs). The used data points have been randomly selected from the time points in the lower left corner. We point out that in Figure 2 and 3 such additional area selection was not applied. We have calculated the LH2 spectrum from different degrees of sparseness using the traditional Discrete FT technique, the LASSO, and SEMA. Here we only show compression levels leading to successful reconstructions of experimental 2D spectra. In SI more examples are presented including also the clear failures.

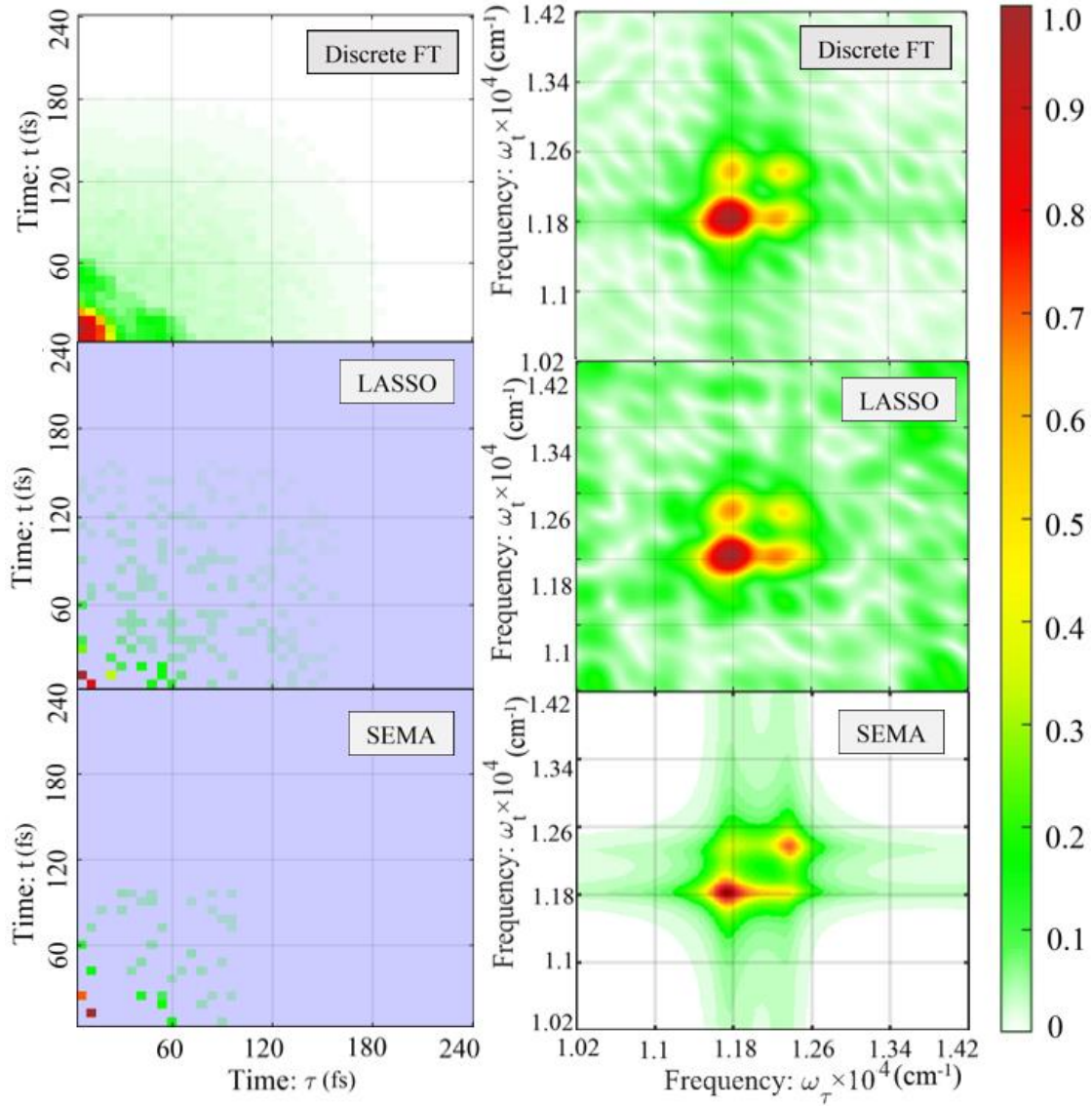

**Figure 4.** The absolute value of the time-domain signal, as well as the Fourier method, the LASSO, and SEMA estimates of the 2D spectral slice at T=70fs. Here, the Fourier method used all the available 40 x 40 = 1600 samples, whereas the LASSO only uses 200 samples (12.5%) and SEMA only 40 samples (2.5%). In case of the latter two, the data were sampled only from the lower left corner of the size 26 x

26 for the LASSO and 16 x 16 for the SEMA. The blue-gray colour shows the data points not used by the LASSO and SEMA.

From the comparison of the simulated and experimental spectrum, it is clear that SEMA requires less data than the LASSO to obtain the reconstructed multidimensional spectra. In the implementation that we use, SEMA explicitly assumes Lorentzian lineshapes (although the method may be modified to also allow for, e.g., Voigt lineshapes). In condensed phase spectroscopy, spectral lineshapes can be highly nontrivial. Here too, while the spectral positions and linewidths are well reconstructed by SEMA with only a fraction of the data points, the spectral shape is quite different from what a direct FT with full data set provides. This is important to keep in mind if detailed information about the lineshapes is part of the analyses. Another issue of practical importance is the overlapping spectra. Here the experimental spectra contain well separated bands. In case of the simulations, in Figure 2 and 3, the spectral overlap may have taken place in some of the random cases. More thorough analyses of this issue goes, however, beyond the frame of the current study.

Finally, in order to relate our work to other analogous studies in coherent 2D spectroscopy we point out that the first work where the compressed sensing was aplied[36] in this context uses a version of the LASSO algorithm. The projection reconstruction method that has been used for speeding up the data collection[37] concerns selecting preferable sampling points carrying more information than those on a uniform grid. Both LASSO and SEMA methods could then be applied on those samples which would further improve the performance.

## CONCLUSIONS

The power of the LASSO and SEMA estimators in reconstructing coherent 2D spectra were analysed by applying the methods to sparsely sampled model data with known spectral parameters. Both estimates are able to reconstruct the spectra using only a fraction of the full data set, achieving better performance than the traditional Fourier technique. This allows for a drastic reduction of the required measurement time for a given experiment. We also sparsely sampled and reconstructed the experimental coherent multidimensional spectra of the antenna complex LH2. Of the studied estimators, SEMA has been shown to offer preferable estimates, and is the technique that is generally recommended by us. Though, since SEMA explicitly assumes Lorentzian lineshape, it is not suitable if analyses of a general complicated spectral lineshape is needed.

## AUTHOR INFORMATION

**Corresponding Authors**

***E-mail:** khadga_jung.karki@chemphys.lu.se
***E-mail:** andreas.jakobsson@matstat.lu.se
***E-mail:** tonu.pullerits@chemphys.lu.se

**Notes**
The authors declare no competing financial interest.

## ACKNOWLEDGMENTS

We acknowledge financial support from the eSSENCE, the Swedish Research Council, KAW foundation, the Crafoord Foundation, the JRA program of the Laserlab-Europe EU-H2020 654148 grant, and NanoLund. Z. W. is grateful to China Scholarship Council (CSC) for the financial support.

## REFERENCES

(1) Mukamel, S.; Tanimura, Y.; Hamm, P. Coherent Multidimensional Optical Spectroscopy. *Acc. Chem. Res.* **2009**, *42*, 1207-1209.
(2) Brixner, T.; Stenger, J.; Vaswani, H. M.; Cho, M.; Blankenship, R. E.; Fleming, G. Two-dimensional

spectroscopy of electronic couplings in photosynthesis. R. *Nature*. **2005**. *434*, 625-628.

(3) Pullerits, T.; Zigmantas, D.; Sundström, V. Beatings in electronic 2D spectroscopy suggest another role of vibrations in photosynthetic light harvesting. *Proc. Natl. Acad. Sci.* **2013**, *110*, 1148 -1149.

(4) Lenngren, N.; Abdellah, M. A.; Zheng. K.; Al-Marri, M. J.; , D.; Zidek, K.; Pullerits, T. Hot electron and hole dynamics in thiol-capped CdSe quantum dots revealed by 2D electronic spectroscopy. *Phys. Chem. Chem. Phys*. **2016**, *18*, 26199-26204.

(5) Yurs, L. A.; Block, S. B.; Pakoulev, A. V.; Selinsky, R. S.; Jin, S., Wright, J. Multiresonant Coherent Multidimensional Electronic Spectroscopy of Colloidal PbSe Quantum Dots. *J. Phys. Chem. C*. **2011**, *115*, 22833–22844.

(6) Tollerud, J. O.; Cundiff, S. T.; Davis, J. A. Revealing and Characterizing Dark Excitons through Coherent Multidimensional Spectroscopy. *Phys. Rev. Lett.* **2016**, *117*, 97401.

(7) Lim, J.; Palecek, D.; Caycedo-Soler, F.; Lincoln, C. N; Javier Prior, J.; Berlepsch, H. V.; Huelga, S. F.; Plenio, M. B.; Zigmantas, D.; Hauer, J. Vibronic origin of long-lived coherence in an artificial molecular light harvester. *Nat. Commun.* **2015**, *9*, 1-7.

(8) Milot, F.; Prokhorenko, V. I.; Mancal, T.; Berlepsch, H. V.; Bixner, O.; Kauffmann, H. F.; Hauer, J. Vibronic and Vibrational Coherences in Two-Dimensional Electronic Spectra of Supramolecular J-Aggregates. J. Phys. Chem. A. **2013**, *117*, 6007−6014.

(9) Abramavicius, D.; Palmieri, B.; Voronine, D. V; Šanda, F.; Mukamel, S. Coherent Multidimensional Optical Spectroscopy of Excitons in Molecular Aggregates; Quasiparticle versus Supermolecule Perspectives. *Chem. Rev.* **2009**, *109*, 2350–2408.

(10) Brixner, T.; Stiopkin, I. V.; Fleming, G. R. Tunable Two-Dimensional Femtosecond Spectroscopy. *Opt. Lett.* **2004**. *29*, 884-886.

(11) Drager, S.; Roeding, S.; Brixner, T. Rapid-scan coherent 2D fluorescence spectroscopy. *Opt. Express.* **2017**, *25*, 3259-3267.

(12) Karki, K. J.; Widom, J. R.; Seibt, J.; Moody, I.; Lonergan, M. C.; Pullerits, T.; Marcus, A. H. Coherent two-dimensional photocurrent spectroscopy in a PbS quantum dot photocell. *Nat. Commun*. **2014**, 5, 5869.

(13) Aeschlimann, M.; Brixner, T.; Fischer, A.; Kramer, C.; Melchior, P.; Pfeiffer, W.; Schneider, C.; Strüber, C.; Tuchscherer, P.; Voronine, D. V. Coherent Two-dimensional Nanoscopy. *Science*. **2011**, *333*, 1723-1726.

(14) Roeding, S.; Brixner, T. Coherent two-dimensional electronic mass spectrometry, *Nat. Commun*. **2018**, *28*, 1-9.

(15) Tibshirani, R. Regression Shrinkage and Selection via the Lasso. *J. R. Stat. Soc*. **1996**, 58, 267-288.

(16) Donoho, D. L. Compressed sensing. *IEEE T. Inform. Theory*. **2006**, *52*, 1289–1306.

(17) Greenberg, J.; Krishnamurthy, K.; Brady, D. Compressive single-pixel snapshot x-ray diffraction imaging. *Opt. Lett.* **2014**, *39*, 111–114.

(18) Dunbar, J. A.; Osborne, D. G.; Anna, J. M.; Kubarych, K. J. Accelerated 2D-IR Using Compressed Sensing. *J. Phys. Chem. Lett.* **2013**, *4*, 2489–2492.

(19) Bostock, M. J.; Holland, D. J.; Nietlispach, D. Compressed sensing reconstruction of undersampled 3D NOESY spectra: application to large membrane proteins. *J. Biomol. NMR*. **2012**, *54*, 15–32.

(20) Luo, Y. L.; Andersson, S. B. A comparison of reconstruction methods for undersampled atomic force microscopy images. *Nanotechnology*, **2015**, *26*, 1-12.

(21) Chen, E. X.; Russell, Z. E.; Amsden, J. J.; Wolter, S. D.; Danell, R. M.; Parker, C. B.; Brady, D. J. Order of Magnitude Signal Gain in Magnetic Sector Mass Spectrometry Via Aperture Coding. *J. Am. Soc. Mass Spectrom*. **2015**, *26*, 1633–1640.

(22) Spencer, A. P.; Spokoyny, B.; Ray, S.; Sarvari, F.; Harel, E. Mapping multidimensional electronic structure and ultrafast dynamics with single-element detection and compressive sensing, *Nat. Commun*. **2016,** *7*, 1-6.

(23) Swärd, J.; Elvander, F.; Jakobsson, A. Designing sampling schemes for multi-dimensional data. *Signal Process.* **2018**, *150*, 1–10.

(24) Swärd, J.; Adalbjörnsson, S. I.; Jakobsson, A. High resolution sparse estimation of exponentially decaying N-dimensional signals. *Signal Process*. **2016**, *128*, 309–317.

(25) Sundström, V.; Pullerits, T.; van Grondelle, R. Photosynthetic Light-Harvesting: Reconciling Dynamics and Structure of Purple Bacterial LH2 Reveals Function of Photosynthetic Unit. *J. Phys. Chem. B*. **1999**, *103*, 2327–2346.

(26) Yoneda, Y.; Noji, T.; Katayama, T.; Mizutani, N.; Komori, D.; Nango, M.; Dewa, T. Extension of Light-Harvesting Ability of Photosynthetic Light-Harvesting Complex 2 (LH2) through Ultrafast Energy Transfer from Covalently Attached Artificial Chromophores. *J. Am. Chem. Soc.* **2015**, *137*, 13121–13129.

(27) Harel, E.; Engel, G. S. Quantum coherence spectroscopy reveals complex dynamics in bacterial light-harvesting complex 2 (LH2). *Proc. Natl. Acad. Sci*. **2012**, *109*, 706-711.

(28) McDermott, G.; Prince, S. M.; Freer, A. A.; Hawthornthwaite-Lawless, A. M.; Papiz, M. Z.; Cogdell, R. J.; Isaacs, N. W. Crystal structure of an integral membrane light-harvesting complex from photosynthetic bacteria. *Nature.* **1995**, *374*, *517-521*.

(29) Bandilla, M.; Ücker, B.; Ram, M.; Simonin, I.; Gelhaye, E.; McDermott, G.; Scheer, H. Reconstitution of the B800 bacteriochlorophylls in the peripheral light harvesting complex B800–850 of Rhodobacter sphaeroides 2.4.1 with BChl a and modified (bacterio-)chlorophylls. *Biochim. Biophys. Acta – Bioenerg*. **1998**, *1364*, 390–402.

(30) Schröter, M.; Pullerits, T.; Kuhn, O. Using fluorescence detected two-dimensional spectroscopy to investigate initial exciton delocalization between coupled chromophores, *J. Chem Phys.* **2018**, *149*, 114107.

(31) Tiwari, V.; Matutes, Y. A.; Gardiner, A. T.; Jansen, T. L. C.; Cogdell, R. J.; Ogilvie, J. P. Spatially-resolved fluorescence-detected two-dimensional electronic spectroscopy probes varying excitonic structure in photosynthetic bacteria, *Nat. Commun*. **2018,** *9*, 4218.

(32) Karki, K. J.; Chen, J. S.; Sakurai, A.; Shi, Q.; Gardiner, A. T.; Kuhn, O.; Cogdell, R. J.; Pullerits, T. Before Förster. Initial excitation in photosynthetic light harvesting. *Chem. Sci.* **2019**, *10*, 7923-7928.

(33) Tekavec, P. F.; Lott, G. A. Fluorescence-detected two-dimensional electronic coherence spectroscopy by acousto-optic phase modulation. *J. Chem. Phys.* **2007**, *127*, 1-21.

(34) Fu, S. Y.; Sakurai, A.; Liu, L.; Edman, F.; Pullerits, T.; Öwall, V.; Karki, K. J. Generalized lock-in amplifier for precision measurement of high frequency signals. *Rev. Sci. Instrum*. **2013**, *84*, 1-5.

(35) Roeding, S.; Klimovich, N.; Brixner, T. Optimizing sparse sampling for 2D electronic spectroscopy. *J. Chem. Phys*. **2017**, *146*, 84201.

(36) Sanders, J. N.; Saikin, S. K.; Mostame, S.; Andrade, X.; Widom, J. R.; Marcus, A. H.; Aspuru-Guzik, A. Compressed Sensing for Multidimensional Spectroscopy Experiments. *The J. Phys. Chem. Lett*. **2012**, *3*, 2697–2702.

(37) Hutson, W. O.; Spencer, A. P.; Harel, E. Ultrafast Four-Dimensional Coherent Spectroscopy by Projection Reconstruction. *The J. Phys. Chem. Lett*. **2018**, *9*, 1034–1040.

(38) Elvander, F.; Swärd, j.; Jakobsson, A. Mismatched Estimation of Polynomially Damped Signals. *IEEE International Workshop on Computational Advances in Multi-Sensor Adaptive Processing*. **2019**. 15-18.